\documentclass[final]{aipproc}

\usepackage{amssymb}
\usepackage{amsmath}

\layoutstyle{6x9}

\newcommand{\zp}{Z^{\prime}}
\newcommand{\mzp}{M_{\tilde{Z}^\prime}}
\newcommand{\uonep}{\ensuremath{U(1)^\prime}}
\newcommand{\gz}{g_{z^\prime}}

\begin{document}

%preprint number

\rightline{EFI 09-27}\mbox{}\\
\vspace{-4em}

\title{From A to  $\mathbf \zp$: Combining Anomaly and $\mathbf \zp$ Mediation
of Supersymmetry Breaking}

\classification{12.60.Jv, 14.80.Ly}
\keywords      {$\zp$ mediation, Anomaly mediation}

\author{Gil Paz}{
  address={ Enrico Fermi Institute, University of Chicago, 5640
S. Ellis Ave., Chicago, IL 60637, U.S.A} }

\begin{abstract}
Combining anomaly with $\zp$ mediation allows us to solve the
tachyonic slepton problem of the former and avoid fine tuning in the
latter. We describe how the two mechanisms can be combined, and some
of the phenomenology of such a joint scenario.
\end{abstract}

\maketitle

\section{Introduction}

$\zp$ mediation of supersymmetry (SUSY) breaking is a mediation
mechanism in which both the hidden and the visible sectors are charged
under a new $\uonep$ gauge interaction. Such a possibility is
motivated by the fact that many superstring constructions contain such
an ``extra'' $\uonep$ (see references in
\cite{Langacker:2008yv}). This interaction can then mediate SUSY
breaking to the visible sector.  In this scenario, one assumes that
the gauge interaction is unbroken in the hidden sector, and at a scale
$\Lambda_S$ SUSY is broken and a $\zp$ gaugino mass, $\mzp$, is
generated. Since all the visible sector chiral superfields are charged
under this $\uonep$, the scalars receive a mass at one loop order,
while the $SU(3)_C \times SU(2)_L \times U(1)_Y$ (``MSSM'') gauginos
get a mass only at two loop order. As a result, the soft scalar masses
are about 1000 times heavier than the gaugino masses. Since direct
searches constrain the gaugino masses to be above 100 GeV, we have
basically two options.

The first is to assume that the gaugino masses are around 100-1000
GeV. The soft scalar masses are then of the order of 100-1000 TeV and
in order to obtain electroweak symmetry breaking at its observed
scale, one fine tuning is needed. Such an approach was explored in the
original $\zp$ mediation papers \cite{ZpSB}.  The second option is to
assume that the scalar masses are around 100-1000 GeV. The ``MSSM''
gauginos are then too light, and they must receive additional
contributions from another mediation mechanism, for example anomaly
mediation (AMSB) \cite{AMSB}.

One can motivate combining anomaly and $\zp$ mediation in the
following heuristic way. For AMSB the soft scalar masses squared can
be ``so small'' that they can actually be negative. For $\zp$ mediation,
on the other hand, they are ``too big'' compared to the electroweak
scale (squared). By combining the two mechanisms we can hope to obtain
scalar masses which are ``just right'', i.e. of the order of the
electroweak scale. Another motivation is that they both naturally
arise from an extra dimensional model, as we show in the next section.

In order for such a combined scenario to be viable, we must demand
that the AMSB contribution to the soft scalar mass squared, which is
roughly $m_{3/2}^2/(16\pi^2)^2$, is comparable to the $\zp$
mediation contribution, which is roughly $\mzp^2/16\pi^2$. Here
$m_{3/2}$ is the gravitino mass.  This implies that $m_{3/2}$ should
be about an order of magnitude larger than $\mzp$. If such a hierarchy
holds, the $\zp$ contribution to the MSSM gaugino masses is three
orders of magnitude suppressed compared to the anomaly contribution
and therefore completely negligible. We will now show that such a mild
hierarchy is natural within an extra dimensional model.

\section{``$\mathbf\zp$-gaugino mediation''}

In the original $\zp$ mediation papers \cite{ZpSB} the mechanism under
which the $\zp$-gaugino becomes massive was left unspecified. Here we
consider a specific implementation that can be thought of as
``$\zp$-gaugino mediation''. As in the gaugino mediation scenario
\cite{Kaplan:1999ac,Chacko:1999mi}, we assume that SUSY is broken on a
spatially separated brane and as a result a gaugino mass term is
generated. Unlike the standard gaugino mediation, we assume that
\emph{only} the $\zp$ gaugino mass is generated, while the ``MSSM''
gauginos remain massless. For example, if we consider one extra
dimension, we can have a brane localized term of the form
\begin{equation}\label{eq:brane}
 c\,\int d^2\theta\, \frac{X}{M_*^2}\,W_{z'}\,W_{z'}\delta(y-L),
\end{equation}
where $W_{z'}$ is the $\uonep$ field strength, $X$ is the field whose
$F$ component generates the gaugino mass, $L$ is the size of the extra
dimension, $M_*$ is the 5D Planck mass and $c$ is a constant. The
relation between the 5D and the 4D Planck mass is $M_*^3\,L=M_P^2$.

When the field $X$ develops an $F$ term, a $\zp$ gaugino mass is
generated:
\begin{equation}
\mzp=c\,\frac{F_X}{M_*^2\,L},
\end{equation}
where the extra factor of $L$ arises from different 4D and 5D
normalizations.  The gravitino mass is of the order
\begin{equation}
m_{3/2}\sim\frac{F}{M_p}=\frac{F}{\sqrt{M_*^{3}\,L}}.
\end{equation}
If we assume that $F \sim F_X$ and define $r \equiv m_{3/2}/\mzp$,
then we should demand that $M_*L\sim c^2r^2$. This product of the 5D
Planck mass and the size of the extra dimension is bounded both from
above and below.  First, to ensure that the gauge coupling are
perturbative \cite{Kaplan:1999ac}, we need  $M_*L\lesssim 16\pi^2$.
Second, to suppress contact terms of the form
\begin{equation}
\frac1{M_*^2}\int d^4\theta\,Y^\dagger\,Y\,Q^\dagger\,Q,
\end{equation}
with $Y (Q)$ hidden (visible) sector fields, which can potentially
violate flavor constraints \cite{Kaplan:2000jz}, we need $M_*L\gtrsim
16$. These two conditions imply that
\begin{equation}\label{eq:allowed}
4\lesssim c\, r\lesssim 4\pi.
\end{equation}
With an order one coefficient in equation (\ref{eq:brane}) we can
easily generate the appropriate hierarchy. We emphasize that the
``$\zp$-gaugino mediation'' scenario we have presented is a special
case of the more general $\zp$ mediation and the two are not
equivalent.

\section{Specific Implementation}
We now present an explicit implementation of the combined scenario. We
choose to do that using the same model for which the original $\zp$
mediation mechanism was implemented \cite{ZpSB}.  One interesting
feature of this model is that the beta function of the strong coupling
vanishes at one loop order. This is not an accident, but a rather
general result following from $SU(3)^2_C\times\uonep$ anomaly
cancellation condition and very general assumptions \cite{ZpSB}. As a
result the gluino mass receives non-zero contributions only at two
loop order, and for a generic choice of parameters, the gaugino mass
hierarchy is $M_1>M_3>M_2$. This should be compared to the ``standard''
AMSB for which the gluino is heavier than the bino and the wino.

To show that our model can lead to a reasonable spectrum we choose one
specific illustration point. We list the input parameters and the
resulting spectrum. The dimensionful input parameters are $m_{3/2}$,
$\mzp$ and $\Lambda_S$:
\begin{equation}
m_{3/2}=80\, {\rm TeV}, \quad \mzp=15\, {\rm TeV}, \quad \Lambda_S\sim
10^6\, {\rm TeV}.
\end{equation}
The ratio of the gravitino mass to the $\zp$-gaugino mass is within
the allowed range of equation (\ref{eq:allowed}).  The dimensionless
input parameters are the $U(1)'$ charges of $H_u$ and $Q$ (the quark
doublet), the $U(1)'$ gauge coupling $\gz$, and the superpotential
couplings $y_t,y_b,y_\tau,y_D,y_E$ and $\lambda$. We take
\begin{eqnarray}
\uonep \mbox{ charges}&:&\quad Q_{H_u}=-\frac25,\quad
Q_Q=-\frac13\nonumber\\ 
\uonep \mbox{ gauge coupling (at }
\Lambda_{S})&:&\quad \gz=0.45\nonumber\\ 
\mbox{Superpotential
  parameters (at } \Lambda_{\rm EW}) &:&\quad\lambda=0.1,\, y_D=0.3,\,
y_E=0.5,\nonumber\\ 
&&\quad y_t = 1, \, y_b = 0.5,\,y_\tau =0.294.
\end{eqnarray}
The values of $y_t,\,y_b$, and $y_\tau$ are chosen to reproduce the
values of the top, the bottom, and the tau masses at the electroweak
scale.

Running down to the electroweak scale we find the following vacuum parameters:
\begin{equation}
\tan\beta=29,\quad \langle S\rangle=11.9 \,{\rm TeV}.
\end{equation}
Due to the large $\tan\beta$ and the large vacuum expectation value
(vev) of $S$, the vevs are strongly ordered : $\langle H_d^0\rangle
\ll\langle H_u^0\rangle\ll\langle S\rangle$.  As a result there is
very little mixing in the extended Higgs and neutralino sectors. Here
we highlight some of the important details of the spectrum. The full
description of the spectrum appears in \cite{deBlas:2009}.
\begin{itemize}
\item ``Higgs'' particles {\bf including one loop radiative
corrections}
$$ m_{h^0}=0.138 \,{\rm TeV},\quad m_{H_1^0}=2.79\,{\rm TeV} ,\quad
m_{H_2^0}=4.78\,{\rm TeV} $$
\item Neutralinos :
$$m_{\tilde{N}_1}=0.278 \,{\rm TeV\, (``Wino")}, \quad
  m_{\tilde{N}_2}=0.61 \,{\rm TeV\, (``Singlino")}, \quad
  m_{\tilde{N}_3}=1.15 \,{\rm TeV\, (``Bino")}$$
$$m_{\tilde{N}_4}\sim m_{\tilde{N}_5}\sim 1.2 \,{\rm TeV\, (``Higgsinos")}
  ,\quad m_{\tilde{N}_6}=12.7 \,{\rm TeV\, (``Z' gaugino")} $$
\item Charginos
$$m_{\tilde{C}_1}=0.278 \,{\rm TeV\, (``Wino")} ,\quad
m_{\tilde{C}_2}=1.2 \,{\rm TeV\, (``Higgsino")}$$
\item Gluino
$$\quad M_3=0.4 \,{\rm TeV}$$
\item Z' gauge boson 
$$M_{Z'}=2.78 \,{\rm TeV}$$
\item MSSM sfermions
$${\rm Lightest}:\, m_{\tilde{b}_1}\sim m_{\tilde{t}_1}=0.7\,{\rm
  TeV},\quad {\rm Heaviest}:\, m_{\tilde{e}_R}\sim
  m_{\tilde{\mu}_R}=12.2 \,{\rm TeV}$$
\item Exotic sfermions
$${\rm Lightest}:\, m_{\tilde{D}_1}=2.53 \,{\rm TeV},\quad {\rm
Heaviest}:\, m_{\tilde{E}_2}=12.8 \,{\rm TeV}$$
\item Exotic fermions
$$m_{D}=3.57 \,{\rm TeV},\quad m_{E}=5.95 \,{\rm TeV}.$$
\end{itemize}

\section{conclusions} 
Combining $\zp$ and anomaly mediation allows us to avoid fine tuning
from the $\zp$ side, and the tachyonic slepton problem from the
anomaly side. It requires a mild mass hierarchy between the gravitino
and the $\zp$ gaugino, which can be obtained from an extra dimensional
model. We have presented an explicit implementation, in which unlike
``standard'' AMSB, the gluino is lighter than the bino. 
%The gluino can
%then be copiously produced and decay to third generation quarks.
The gauginos, in particular the
gluino, are typically lighter than the
sfermions. 
The spectrum also includes a 2.8 TeV $Z'$. A more
detailed description 
of this combined scenario 
will appear in
\cite{deBlas:2009}.

\begin{theacknowledgments}
This work is supported in part by the Department of Energy grants
DE-FG02-90ER40542 and DE-FG02-90ER40560, and by the United
States-Israel Bi-national Science Foundation grant 2006280.
\end{theacknowledgments}

\end{document}